\documentclass[graybox]{svmult}
\usepackage{astrojournals}
\usepackage{natbib}
\usepackage{url}

\usepackage{mathptmx}
\usepackage{helvet}
\usepackage{courier}
\usepackage{type1cm}

\usepackage{makeidx}
\usepackage{graphicx}

\usepackage{multicol}
\usepackage[bottom]{footmisc}

\usepackage{bm}

\usepackage[pdfpagemode={UseOutlines},bookmarks,bookmarksopen,colorlinks,linkcolor={black},citecolor={black},urlcolor={black},breaklinks=true]{hyperref}
\usepackage{amssymb}

\def \mathbi#1{\textbf{\em #1}}
\def \d{\mathrm{d}}

\def\website#1{web: \url{#1}}

\makeindex

\begin{document}

\title*{Warp propagation in astrophysical discs}
\titlerunning{Warp propagation} 
\author{Chris~Nixon and Andrew King}
\authorrunning{Nixon \& King} 
\institute{Chris Nixon \at JILA, University of Colorado \& NIST, Boulder CO 80309-0440, USA\\Einstein Fellow\\\email{chris.nixon@jila.colorado.edu}\\\website{http://jila.colorado.edu/chrisnixon/}
\and Andrew King \at Department of Physics and Astronomy, University of Leicester, Leicester LE1 7RH, UK\\\email{ark@astro.le.ac.uk}\\\website{http://www2.le.ac.uk/departments/physics/people/andrewking/ark}}
\maketitle

\abstract*{Astrophysical discs are often warped, that is, their orbital planes change with radius. This occurs whenever there is a non--axisymmetric force acting on the disc, for example the Lense--Thirring precession induced by a misaligned spinning black hole, or the gravitational pull of a misaligned companion. Such misalignments appear to be generic in astrophysics. The wide range of systems that can harbour warped discs -- protostars, X--ray binaries, tidal disruption events, quasars and others -- allows for a rich variety in the disc's response. Here we review the basic physics of warped discs and its implications.}

\abstract{Astrophysical discs are often warped, that is, their orbital planes change with radius. This occurs whenever there is a non--axisymmetric force acting on the disc, for example the Lense--Thirring precession induced by a misaligned spinning black hole, or the gravitational pull of a misaligned companion. Such misalignments appear to be generic in astrophysics. The wide range of systems that can harbour warped discs -- protostars, X--ray binaries, tidal disruption events, quasars and others -- allows for a rich variety in the disc's response. Here we review the basic physics of warped discs and its implications.\vspace{1.5in}}

\hspace{-0.18in}{\it Big whorls have little whorls, which feed on their velocity. And little whorls have lesser whorls, and so on to viscosity.}\vspace{0.1in}\\\phantom{~~~~~}-- Adaptation of \href{http://en.wikipedia.org/wiki/The_Siphonaptera}{{\it The Siphonaptera}} by Lewis Richardson

\clearpage

\section{Introduction}
\label{intro}
In general gas possesses angular momentum and orbits any central massive object such as a protostar or a black hole rather than falling directly on to it. The test particle orbits are not closed ellipses but form rosettes, as perturbations such as nearby stars or gas self--gravity mean that the potential is not perfectly Keplerian. These orbits intersect and so the gas shocks, causing dissipation of orbital energy through heating and radiation. However angular momentum must be conserved during this process. Therefore the gas settles into the orbit of lowest energy for the given angular momentum -- a circle. In most cases the gas does not all have the same specific angular momentum and forms a disc rather than a ring. More generally gas has angular momentum with a spread of directions and settles into a warped disc, where the disc plane changes with radius. An initially planar disc may become warped through a variety of different effects. If the disc is misaligned with respect to a component of the potential these include Lense--Thirring precession from a spinning black hole \citep{BP1975}, gravitational torques from a companion \citep{PT1995} or torques from a misaligned magnetic field \citep{Lai1999}. An initially aligned disc may also be unstable to warping through processes such as tides \citep{Lubow1992a} or radiation from the central object \citep{Pringle1996}.

A planar (unwarped) disc (\citealt{PR1972}; \citealt{Pringle1981}) evolves under the action of a turbulent viscosity. For simplicity this is often parameterised in the Shakura--Sunyaev form where the viscosity is written as $\nu = \alpha c_{\rm s} H$ \citep{SS1973}. The most likely process driving viscosity is the magneto--rotational instability (MRI) suggested by \cite{BH1991}. This instability requires a weakly magnetised disc whose angular velocity decreases outwards, as is the case in Keplerian and near--Keplerian discs. The cartoon picture of the instability has two parcels of gas, at different radii, joined by a magnetic field loop. The inner parcel rotates faster, so the loop becomes stretched. The magnetic field acts to slow the inner parcel and speed up the outer parcel, thus transferring angular momentum from the inner parcel to the outer parcel. The net effect is that the inner parcel moves inwards through the disc and the outer parcel moves outwards. In reality this process drives turbulence in the disc, which then ultimately transports angular momentum.

The Shakura--Sunyaev viscosity parameterisation is a simple but powerful and intuitively appealing way of characterising the angular momentum transport in a planar disc. As described in \cite{Pringle1981} any turbulent eddies must have a lengthscale smaller than the disc thickness $H$. Moreover any supersonic turbulent motions in the disc shock and rapidly dissipate, giving a maximum signal velocity of approximately the sound speed $c_{\rm s}$. So the maximal viscosity expected in a disc is $\nu_{\rm max} = c_{\rm s}H$, and we can write the viscosity with the Shakura--Sunyaev $\alpha$ parameter as $\nu = \alpha c_{\rm s}H$, but it is important to realise that $\alpha$ need not be constant in position or time. For fully ionised discs, observations suggest that $\alpha \approx 0.1- 0.4$ \citep{Kingetal2007}. Numerical simulations of the MRI without net vertical field do not yet reproduce this result, instead finding $\alpha \approx {\rm a~few}\times 10^{-2}$ \citep[e.g.][]{Simonetal2012}. In other systems where the gas is neutral or only partially ionised, observations allow the MRI--driven $\alpha$ to be lower. For example, observations of protostellar discs find $\alpha \approx 0.01$ \citep[e.g.][]{Hartmannetal1998}, but in dead zones this can be much lower ($\sim 10^{-4}$, e.g. \citealt{FS2003}; \citealt{Simonetal2011}; \citealt{Gresseletal2012}; \citealt{ML2014}).

In a planar disc, where the gas is essentially hydrostatic in the vertical direction, the evolution is usually described by a one--dimensional viscosity which communicates angular momentum radially. However, in a warped disc the communication of angular momentum is three--dimensional. Early investigations (e.g. \citealt{BP1975}; \citealt{Petterson1977a}; \citealt{Petterson1978}; \citealt{Hatchettetal1981}) simply assumed that the viscosity could be described by the same $\nu$ in all directions, and the evolution equations they derived did not conserve angular momentum. 

The first self--consistent investigation was provided by \cite{PP1983}. They give two derivations of the equations of motion for a warped disc. The first (their Section~2) is close to the previous derivations in the literature, but gives the correct form of the internal torques required to conserve angular momentum. The second derivation (their Section~3) is the first to take the internal fluid dynamics into account. This derivation is the first to discuss the radial pressure gradients induced by the warp that oscillate at the orbital frequency, creating a resonance. Therefore the torque corresponding to the vertical shear (or more precisely the radial communication of the component of angular momentum parallel to the local orbital plane) depends {\it inversely}\footnote{See the definition of the quantity $A$ at the end of page 1189 of \cite{PP1983}.} on the viscosity parameter, $\alpha$, as the resonant forcing drives shearing motions which are damped by $\alpha$ \citep[see Section~4.1 of][for a discussion]{LP2007}. An implication of \cite{PP1983} is that simple approaches to the dynamics of warped discs can miss subtle but important effects. But the complex nature of warped discs makes both simple and elaborate approaches necessary -- for physical insight and to get the physics right. We shall see that the simple approach can capture most, but not all, of the relevant physics.

The work of \cite{PP1983} formally requires $H/R < \alpha \ll 1$, and most restrictively, that the disc tilt ($U_z/R\Omega$ in their notation, where $U_z$ is the perturbed vertical velocity) be $\ll H/R$. This last restriction requires that the warp be so small as to be unobservable. So to make progress, \cite{Pringle1992} derived an equation of motion using only conservation laws, and not the internal hydrodynamics, but retaining full generality in terms of the disc tilt. It is this approach that we will follow in deriving the evolution equations below (Section~\ref{diffeqns}). These equations are formally valid (in that they conserve angular momentum) for arbitrary disc warps and viscosities ($\nu_1$,$\nu_2$). However, this approach offers no insight into determining the values of the viscosities or how they depend on any system parameters such as warp amplitude. In this sense its treatment of the internal fluid dynamics is ``naive'' \citep{PP1983}.

For this reason, \cite{Ogilvie1999} started from fully three--dimensional fluid equations (compressible, with a locally isotropic [`Navier--Stokes'] viscosity), and derived a full evolution equation for a warped disc with these properties. This derivation confirmed the equations derived by Pringle, with two important differences. First, the Pringle equations did not include a torque that tends to cause rings to precess when tilted with respect to their neighbours, and second, the torque coefficients are determined as a function of warp amplitude and $\alpha$. The extra torque between rings is not required for angular momentum conservation, so it does not appear in the derivation in \cite{Pringle1992} and the torque coefficients are determined by the local fluid dynamics between rings. \cite{PP1983} derived the torque coefficients for the azimuthal and vertical shear terms in the linear approximation, obtaining the well-known result that  $\alpha_1 = \alpha$ \& $\alpha_2 = 1/\left(2\alpha\right)$. \cite{Ogilvie1999} extends this to the fully nonlinear regime, determining the coefficients for arbitrary warp amplitudes. This work marked a great advance in the understanding of warped discs. \cite{Ogilvie2000} extends these equations to include the effects of viscous dissipation and radiative transport.

In cases where $\alpha < H/R$ the evolution of a warped disc is not controlled by viscosity. Pressure forces dominate and can propagate warping disturbances through distances $> R$ as waves. This occurs because the $\alpha$ damping is too slow to damp the wave locally (on scales of order $H$) allowing it to propagate. A second, more subtle, condition must also be satisfied for wave propagation to be efficient -- the disc must be close to Keplerian, i.e. $\left|1-\kappa^2/\Omega^2\right| \lesssim H/R$ \cite[see e.g. Section 5.1 of][]{WP1999}. This is required to make the forcing resonant in a warp (Keplerian implies the orbital, vertical and epicyclic frequencies are the same, $\Omega = \Omega_z = \kappa$). The first dedicated study of the behaviour of warped discs in this wave--like regime is \cite{PL1995}. This paper showed that these waves, driven by pressure gradients, propagate at approximately half the local sound speed. Further, the propagation behaviour becomes diffusive when a small viscosity $\alpha \sim H/R$ is included. \cite{LP1993} and \cite{KP1995} investigated the propagation of a more general class of waves in discs. As discussed by \cite{Pringle1999} it is possible to recover the warp wave propagation velocity $V_{\rm w} = c_{\rm s}/2$ (for $\alpha=0$) from this approach.

In this review we discuss the two types of warp propagation, through waves and diffusion. We discuss the evolution equations, their interpretation and derivation. We describe the viscosity, and in particular the relation between the small scale turbulent $\alpha$ viscosity and its role in shaping the effective viscosities which control the dynamics of warped discs. Finally we discuss some major results and some outstanding problems in understanding this complex and subtle accretion disc behaviour.

\section{Warp propagation}
\label{prop}
In a planar accretion disc there is only azimuthal shear acting in the disc driving a turbulent viscosity which transports angular momentum. In a warped disc there is a second `vertical' type of shear and so the rate of orbital shear is not simply parallel to the local disc normal $\mathbi{l}$, but is given by \citep[e.g.][]{OL2013a}
\begin{equation}
\mathbi{S} = R\frac{\partial \mathbi{s}}{\partial R} = R\frac{{\rm d}\Omega}{{\rm d} R}\mathbi{l} + R\Omega \frac{\partial \mathbi{l}}{\partial R}\,,
\end{equation}
where $\mathbi{s}\left(R,t\right) = \Omega(R)\mathbi{l}\left(R,t\right)$ is the local orbital angular velocity. We can think of the evolution of a warped disc as controlled by two torques, communicating angular momentum along and normal to the local orbital plane respectively.

\begin{figure}
  \begin{center}
    \includegraphics[angle=0,width=\columnwidth]{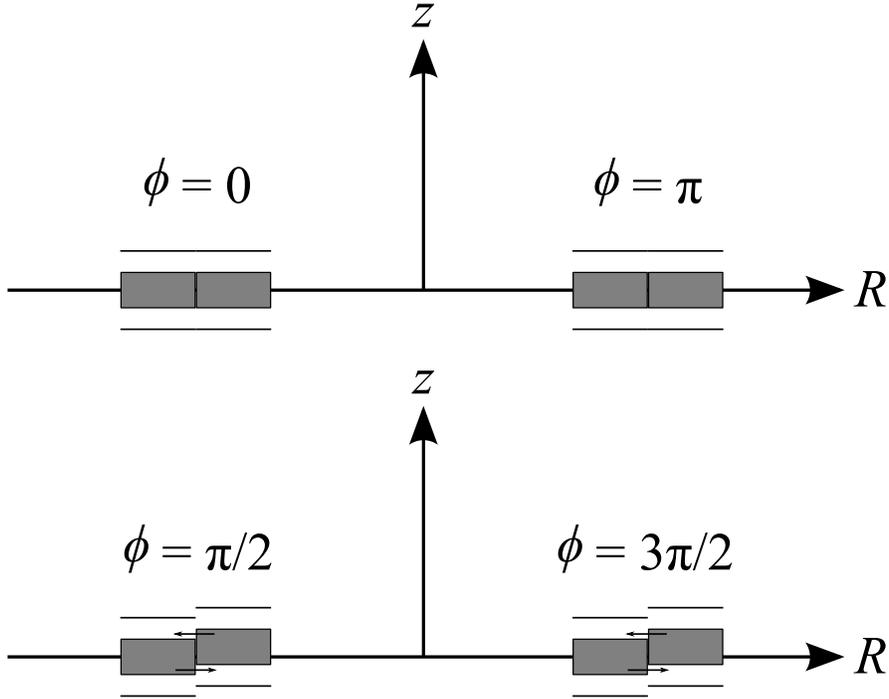}
    \caption{This figure (cf Fig.~10 of \citealt{LP2007}) illustrates the radial pressure gradient induced by a warp. The top and bottom panel show cross-sections of the same two neighbouring  rings of gas, but at different azimuths $\phi$. The shaded regions indicate the higher pressure around the local midplane, and the arrows show the resultant pressure gradient when the rings are misaligned. The azimuthal angle around a ring is measured in the direction of the flow from the descending node where $\phi=0$. The tilted rings cross at the nodes, and so at these points are in aligned contact as usual. At all other azimuths the ring midplanes do not fully line up, causing a region of overpressure above or below the midplane. Each gas parcel feels an oscillating pressure gradient as it orbits in the warp.}
    \label{radpresgrad}
  \end{center}
\end{figure}

The usual planar disc viscosity is driven by azimuthal shear and is assumed to result in a turbulent 
$\alpha$ viscosity acting against the shear and transporting angular momentum radially. The second viscosity associated with the warp is induced by radial pressure gradients driven by the misalignment of neighbouring rings (see Fig.~\ref{radpresgrad}). In a near-Keplerian disc the forcing frequency (epicyclic) and the orbital frequency resonate to produce a significantly enhanced torque, whose response is controlled by $\alpha$. This torque is driven by pressure and results in the launch of a pressure wave. When $\alpha > H/R$ this is damped locally and the evolution is diffusive, but when $\alpha < H/R$ the wave can propagate large distances. This leads to two types of warp propagation in discs; diffusive and wave-like. These are the subject of the next two sections.

The convenient coordinates for describing a warped disc are a hybrid of cylindrical polars and Euler angles. A local annulus (of width $\sim H$) is described by cylindrical polars $\left(R,\phi,z\right)$, but each ring of the disc is tilted in three dimensional space described by the (spherical) Euler angles $\beta\left(R,t\right)$ \& $\gamma\left(R,t\right)$, which correspond to the local disc tilt and twist respectively. So we define the disc unit tilt vector as
\begin{equation}
\label{utv}
\mathbi{l} = \left(\cos\gamma\sin\beta,\sin\gamma\sin\beta,\cos\beta\right)\,.
\end{equation}
The local angular momentum vector for the disc is $\mathbi{L} = \Sigma R^2\Omega\mathbi{l}$, where $\Sigma$ is the disc surface density. The (Cartesian) surface of the disc is 
\begin{equation}
\mathbi{x}\left(R,\phi\right) = R\left(\cos\phi\sin\gamma+\sin\phi\cos\gamma\cos\beta,\sin\phi\sin\gamma\cos\beta-\cos\phi\cos\gamma,-\sin\phi\sin\beta\right)\,,
\end{equation}
where the azimuthal angle $\phi $  is zero at the descending node and increases in the direction of the flow. Further quantities such as the elements of surface area then follow as described in \cite{Pringle1996}. Vector calculus in these warped disc coordinates is comprehensively summarised in Section 2 of \cite{Ogilvie1999}. We next derive the equations of motion for warped accretion discs.

\subsection{Diffusion}
\label{diff}

Here we consider diffusive discs with $H/R < \alpha < 1$. We derive the equations of motion in a simple way, allowing significant insight into the problem. This derivation is naive to the internal fluid dynamics in a warped disc \citep{PP1983}, and so we compare and contrast with the results of more complete derivations that are too involved for our purposes here \citep[e.g.][]{Ogilvie1999}.

\subsubsection{Evolution equations}
\label{diffeqns}
The derivation here is a three--dimensional version of the standard planar accretion disc equations (e.g. \citealt{Pringle1981}; \citealt{Franketal2002}), so we briefly recall the planar case first.

We assume the disc is planar and that all quantities can be azimuthally averaged (requiring that the radial velocity is much smaller than the orbital velocity, $V_R \ll V_\phi$) so all quantities are functions of radius $R$ and time $t$ only. The disc has surface density $\Sigma\left(R,t\right)$, radial velocity $V_R\left(R,t\right)$, and angular velocity $\Omega\left(R,t\right)$. We consider an annulus of gas between $R$ and $R+\Delta R$. The mass in this annulus is $2 \pi R \Delta R \Sigma$. Conservation of mass relates the rate of change of this mass to the net flow of mass into or out of the annulus
\begin{equation}
\frac{\partial}{\partial t}\left(2\pi R\Delta R \Sigma\right) = \left(2\pi R V_R\Sigma\right)_{R} - \left(2\pi RV_R\Sigma\right)_{R+\Delta R}\,.
\end{equation}
Rearranging and taking the limit as $\Delta R \rightarrow 0$ gives the continuity equation (mass conservation) for a disc
\begin{equation}
\label{conmass}
\frac{\partial \Sigma}{\partial t} + \frac{1}{R}\frac{\partial}{\partial R}\left(R\Sigma V_R \right) = 0\,.
\end{equation}
Similarly we can derive the equation expressing angular momentum conservation. The angular momentum of an annulus is $2\pi R \Delta R \Sigma R^2 \Omega$ and the rate of change of angular momentum is the net flux of angular momentum plus the net torques:
\begin{eqnarray}
\frac{\partial}{\partial t}\left(2\pi R \Delta R \Sigma R^2 \Omega \right) & = & \left(2\pi R \Sigma R^2 \Omega V_R\right)_{R} - \left(2\pi R \Sigma R^2 \Omega V_R\right)_{R+\Delta R}\\ \nonumber & &
 +{} G\left(R+\Delta R\right) - G\left(R\right)\,.
\end{eqnarray}
Again by rearranging and taking the limit $\Delta R \rightarrow 0$ we have the angular momentum equation
\begin{equation}
\frac{\partial}{\partial t}\left(\Sigma R^2 \Omega \right) + \frac{1}{R}\frac{\partial}{\partial R}\left(R\Sigma V_R R^2 \Omega \right) = \frac{1}{2\pi R}\frac{\partial G}{\partial R}\,,
\end{equation}
where $G\left(R,t\right)$ is the internal disc torque resulting from the disc viscosity. The viscous force is proportional to the rate of shearing
\begin{equation}
F = 2\pi R H \mu R\frac{\d \Omega}{\d R}\,,
\end{equation}
where $R \d \Omega/\d R$ is the rate of shear and $\mu$ is the dynamic viscosity, related to the kinematic viscosity by $\mu = \rho \nu$, where $\rho$ is the density. Using $\rho = \Sigma/H$ the force is
\begin{equation}
F = 2\pi R \nu \Sigma R\frac{\d \Omega}{\d R}\,.
\end{equation}
Note that this force acts in the azimuthal direction, and thus when taking the cross product with the radial vector to generate the torque we see that the torque acts in the correct direction (i.e. on the $z$--component of angular momentum -- we are only considering a planar disc so far). So we can write the internal viscous torque as
\begin{equation}
\label{visctorque}
G = 2\pi R \nu\Sigma R\Omega^\prime R
\end{equation}
and the equation expressing angular momentum conservation as
\begin{equation}
\label{conangmom}
\frac{\partial}{\partial t}\left(\Sigma R^2 \Omega \right) + \frac{1}{R}\frac{\partial}{\partial R}\left(R\Sigma V_R R^2 \Omega\right) = \frac{1}{R}\frac{\partial}{\partial R}\left(\nu\Sigma R^3 \Omega^\prime\right)\,.
\end{equation}

Multiplying Equation~\ref{conmass} by $R^2\Omega$ and then subtracting from Equation~\ref{conangmom} we rearrange for $V_R$ to get
\begin{equation}
V_R = \frac{\frac{\partial}{\partial R} \left(\nu\Sigma R^3 \Omega^\prime\right)}{R\Sigma\frac{\partial}{\partial R}\left(R^2\Omega\right)}\,.
\end{equation}
Substituting this into (\ref{conmass}) gives
\begin{equation}
\frac{\partial \Sigma}{\partial t} = \frac{1}{R}\frac{\partial}{\partial R} \left(\frac{\frac{\partial}{\partial R} \left[\nu\Sigma R^3 \left(-\Omega^\prime\right) \right]}{\frac{\partial}{\partial R} \left(R^2 \Omega\right)}\right)\,,
\end{equation}
which for the Keplerian potential, $\Omega = \sqrt{GM/R^3}$, is
\begin{equation}
\frac{\partial \Sigma}{\partial t} = \frac{3}{R}\frac{\partial}{\partial R} \left[R^{1/2} \frac{\partial}{\partial R} \left(\nu\Sigma R^{1/2} \right)\right]\,.
\end{equation}
This is a diffusion equation, so if we know the viscosity $\nu$ we know how a distribution $\Sigma\left(R,t\right)$ will evolve. So far we have said nothing about the disc viscosity or what drives it -- we will return to this in detail in Section~\ref{viscosity}.

Assuming the disc is in vertical ($z$--direction) hydrostatic balance we derive the vertical structure by equating the relevant components of the pressure force and gravity
\begin{equation}
\frac{1}{\rho}\frac{\partial P}{\partial z} = -\frac{GM}{R^2 + z^2}\frac{z}{\sqrt{R^2+z^2}}\,,
\end{equation}
where the last factor is the geometrical sine term. If the disc is thin we have $\left|z\right| \ll R$, so
\begin{equation}
\frac{1}{\rho}\frac{\partial P}{\partial z} = -\frac{GMz}{R^3}\,.
\end{equation}
Now if we assume an equation of state, e.g. isothermal $P=c_{\rm s}^2\rho$, then
\begin{equation}
c_{\rm s}^2 \frac{\partial \ln \rho}{\partial z} = -\frac{GMz}{R^3}
\end{equation}
and we can integrate to get
\begin{equation}
\rho = \rho_0 \exp\left(\frac{-GMz^2}{2c_{\rm s}^2 R^3}\right) = \rho_0 \exp\left(-\frac{z^2}{2H^2}\right)\,,
\end{equation}
where we have introduced the disc scale--height $H=c_{\rm s}/\Omega$. From this we see that a disc in vertical hydrostatic equilibrium has $H/R = c_{\rm s}/V_\phi$, so for thin discs the flow is supersonic.\\

We can now attempt a similar calculation for a warped disc, where the internal torques act in three dimensions. This calculation draws heavily on \cite{PP1983} and \cite{Pringle1992}. The local angular momentum density is $\mathbi{L}\left(R,t\right) = \Sigma R^2\Omega \mathbi{l}\left(R,t\right)$, where $\mathbi{l}\left(R,t\right)$ is a unit vector in the direction of the ring angular momentum. The mass conservation equation is the same as for the flat disc (\ref{conmass}). Angular momentum conservation is expressed as\vspace{0.2in}
\begin{eqnarray}
\frac{\partial}{\partial t}\left(2\pi R \Sigma R^2 \Omega \Delta R \mathbi{l} \right) & = & ~~~\left( 2\pi R \Sigma R^2 \Omega V_R \mathbi{l} \right)_R\\ \nonumber & &
 -{}\left( 2\pi R \Sigma R^2 \Omega V_R \mathbi{l} \right)_{R+\Delta R}\\ \nonumber & &
 +{} \mathbi{G}\left(R+\Delta R\right) - \mathbi{G}\left(R\right)\,,
\end{eqnarray}
where $\mathbi{G}$ is the three--dimensional internal torque. Again rearranging and taking the limit $\Delta R \rightarrow 0$ we have
\begin{equation}
\label{evo1}
 \frac{\partial}{\partial t}\left(\Sigma R^2\Omega\mathbi{l}\right) + \frac{1}{R}\frac{\partial}{\partial R}\left(\Sigma V_R R^3\Omega\mathbi{l}\right) = \frac{1}{2\pi R}\frac{\partial \mathbi{G}}{\partial R}\,.
\end{equation}

The viscous torque has two obvious components. The $\left(R,\phi\right)$ stress contributes a torque acting in the direction of $\mathbi{l}$ (cf \ref{visctorque})
\begin{equation}
\label{g1}
\mathbi{G}_1 = 2\pi R \nu_1\Sigma R\Omega^\prime R \mathbi{l}\,,
\end{equation}
where $\nu_1$ is the azimuthal shear viscosity.
For two neighbouring rings with $\mathbi{l}$ and $\mathbi{l} + \Delta\mathbi{l}$, the $\left(R,z\right)$ stress acts to communicate $\Delta\mathbi{l}$ between the rings, so the torque acts in the direction $\partial\mathbi{l}/\partial R$. The $\left(R,z\right)$ torque is then 
\begin{equation}
\label{g2}
\mathbi{G}_2 = 2\pi R \frac{1}{2}\nu_2 \Sigma  R^2\Omega \frac{\partial\mathbi{l}}{\partial R}\,,
\end{equation}
where $\nu_2$ is the vertical shear viscosity and the factor of a half comes from integrating $\cos^2\phi$ (cf Fig.~\ref{radpresgrad}) across the ring \citep{PP1983}. There is also a third component \citep{PP1983,Ogilvie1999} which we could write as 
\begin{equation}
\label{g3}
\mathbi{G}_3 = 2\pi R \nu_3 \Sigma R^2 \Omega \mathbi{l}\times\frac{\partial \mathbi{l}}{\partial R}\,.
\end{equation}
But this defines an effective {\it viscosity} $\nu_3$ -- this is not an appealing notation as this term does not lead to diffusive behaviour. Instead this torque causes a ring to precess if it is inclined with respect to its neighbours. This produces dispersive wave--like propagation of the warp \citep{Ogilvie1999}.

We combine (\ref{g1}), (\ref{g2}) \& (\ref{g3}) to give the internal torque \citep[cf. Equation 55 of][]{OL2013a}
\begin{equation}
\mathbi{G} = 2\pi R \Sigma R^2 \Omega \left[ \nu_1 \left(\frac{\Omega^\prime}{\Omega}\right)\mathbi{l} + \frac{1}{2}\nu_2\frac{\partial \mathbi{l}}{\partial R} + \nu_3 \mathbi{l}\times\frac{\partial \mathbi{l}}{\partial R} \right]\,.
\end{equation}
Now, putting this into (\ref{evo1}) we have
\begin{eqnarray}
\label{evo2}
 \frac{\partial}{\partial t}\left(\Sigma R^2\Omega\mathbi{l}\right) + \frac{1}{R}\frac{\partial}{\partial R}\left(\Sigma V_R R^3\Omega\mathbi{l}\right) & = & ~~~\frac{1}{R}\frac{\partial}{\partial R}\left(\nu_1 \Sigma R^3 \Omega^\prime\mathbi{l}\right)\\ \nonumber & &
 +{} \frac{1}{R}\frac{\partial}{\partial R}\left(\frac{1}{2}\nu_2\Sigma R^3\Omega \frac{\partial \mathbi{l}}{\partial R}\right)\\ \nonumber & &
 +{} \frac{1}{R}\frac{\partial}{\partial R}\left(\nu_3 \Sigma R^3 \Omega \mathbi{l}\times\frac{\partial \mathbi{l}}{\partial R}\right)\,,
\end{eqnarray}
which is identical to that found in \cite{Pringle1992} except for the last term.

\citet{Pringle1992} showed that this equation can be combined with (\ref{conmass}) to eliminate $V_R$. This is done by taking the dot product of $\mathbi{l}$ with (\ref{evo2}) and then subtracting $R^2\Omega \times$ Equation (\ref{conmass}) and using $\mathbi{l}\cdot \partial\mathbi{l}/\partial R = 0$ and $\partial/\partial R \left(\mathbi{l}\cdot \partial\mathbi{l}/\partial R \right) = 0 \Rightarrow \mathbi{l}\cdot \partial^2\mathbi{l}/\partial R^2 = -\left|\partial \mathbi{l}/\partial R \right|^2$. This gives \citep[][Equation 2.3]{Pringle1992}
\begin{equation}
V_R = \frac{\frac{\partial}{\partial R}\left(\nu_1\Sigma R^3\Omega^\prime\right) - \frac{1}{2}\nu_2\Sigma R^3 \Omega \left|\frac{\partial\mathbi{l}}{\partial R}\right|^2}{R\Sigma \frac{\partial}{\partial R}\left(R^2\Omega\right)}\,.
\end{equation}
We note that the extra torque derived by \cite{Ogilvie1999} does not have a component in the direction of $\mathbi{l}$ (cf. Equation 123 of \citealt{Ogilvie1999}). This shows its precessional nature: it causes neighbouring rings to change their planes, but does not drive any radial flux of mass or angular momentum.
Finally, substituting $V_R$ into (\ref{evo2}), we get the evolution equation for the disc angular momentum vector $\mathbi{L}$
\begin{eqnarray}
\label{pringledLdt}
 \frac{\partial \mathbi{L}}{\partial t} & = & ~~~\frac{1}{R}
 \frac{\partial }{\partial R} \left\{ \frac{\left(\partial / \partial
   R \right) \left[\nu_{1}\Sigma R^{3}\left(-\Omega^{'} \right)
     \right] }{\Sigma \left( \partial / \partial R \right) \left(R^{2}
   \Omega \right)} \mathbi{L}\right\} \\ \nonumber & &
 +{}\frac{1}{R}\frac{\partial}{\partial R}\left[\frac{1}{2} \nu_{2}R
   \left| \mathbi{L} \right|\frac{\partial \mathbi{l}}{\partial
     R} \right] \\ \nonumber & &
 +{}\frac{1}{R}\frac{\partial}{\partial R}
 \left\{\left[\frac{\frac{1}{2}\nu_{2}R^{3}\Omega \left|\partial
     \mathbi{l} / \partial R \right| ^{2}}{\left( \partial /
     \partial R \right) \left( R^{2} \Omega \right)} +
   \nu_{1}\left(\frac{R \Omega^{'}}{\Omega} \right) \right]
 \mathbi{L}\right\}\\ \nonumber & &
 +{}\frac{1}{R}\frac{\partial}{\partial R}\left(\nu_3 \Sigma R^3 \Omega \mathbi{l}\times\frac{\partial \mathbi{l}}{\partial R}\right)\,.
\end{eqnarray}
We see that this equation evolves both the disc {\it shape} $\mathbi{l}\left(R,t\right)$ and 
{\it surface density} $\Sigma\left(R,t\right)$ through the angular momentum vector $\mathbi{L}\left(R,t\right) = \Sigma R^2\Omega \mathbi{l}$. This equation is almost exactly that derived by \cite{Pringle1992},  with the addition of the last term derived by \cite{Ogilvie1999}. The last term can also be found in the linear hydrodynamic analysis of \cite{PP1983} where the warp diffusion coefficient is complex (see also Equation 2.1 of \citealt{KP1985}).\footnote{See the definition of the quantity $A$ at the end of page 1189 of \cite{PP1983}. In the linear equations it is often convenient to assume the unit tilt vector $\mathbi{l}=\left(l_x,l_y,l_z\right)$ has $l_z\approx 1$ and therefore adopt complex equations for the disc tilt where e.g. $W=l_x+il_y$. In this case a diffusion coefficient with non--zero real and imaginary parts has components in the direction of both $\partial\mathbi{l}/\partial R$ and $\mathbi{l}\times\partial\mathbi{l}/\partial R$.} This term leads to precession in the presence of a warp \citep{Ogilvie1999}, but its coefficient, $\nu_3$, is smaller than $\nu_2$ by a factor $\approx \alpha$. Therefore in time-dependent problems this term is often neglected \citep[e.g. Section 3.4 of][]{LP2007}. However, in the case of an inviscid non--Keplerian disc, this is the only non--zero internal torque \citep[Section 7.2 of][]{Ogilvie1999}.

If the rotation law and the torque coefficients ($\nu_1$,$\nu_2$,$\nu_3$) are specified in terms of the disc quantities, equation (\ref{pringledLdt}) can be solved numerically, conserving angular momentum to machine precision by the method described in \cite{Pringle1992}. This has been used in many papers to explore the evolution of warped discs (e.g. \citealt{Pringle1997}; \citealt{WP1999}; \citealt{LP2006}; \citealt{NK2012}).

The derivation of this equation was somewhat simplistic, as it did not take the internal fluid dynamics into account \citep{PP1983}. So to check its validity \cite{Ogilvie1999} started from the full three--dimensional fluid--dynamical equations and derived the equations of motion (Equations 121 \& 122 in \citealt{Ogilvie1999}). \cite{Ogilvie1999} was able to confirm the equations derived by \cite{Pringle1992}, with two important refinements. First, the \cite{Pringle1992} equations were missing the $\nu_3$ term which makes neighbouring annuli precess if they are tilted with respect to each other (included above), and second, the torque coefficients are uniquely determined by the warp amplitude and $\alpha$.

For comparison we give the evolution equations derived by Ogilvie, written in our notation but retaining the $Q_i$ coefficient form. The mass conservation equation is
\begin{equation}
\frac{\partial \Sigma}{\partial t} + \frac{1}{R}\frac{\partial}{\partial R}\left(R\Sigma V_R\right) = 0
\end{equation}
and angular momentum conservation is
\begin{eqnarray}
\frac{\partial}{\partial t}\left(\Sigma R^2\Omega \mathbi{l}\right) + \frac{1}{R}\frac{\partial}{\partial R}\left(\Sigma V_R R^3 \Omega \mathbi{l}\right) & = & ~~~\frac{1}{R}\frac{\partial}{\partial R}\left(Q_1\Sigma \frac{c_{\rm s}^2}{\Omega} R^2\Omega \mathbi{l}\right)\\ \nonumber & &
 +{} \frac{1}{R}\frac{\partial}{\partial R}\left(Q_2\Sigma \frac{c_{\rm s}^2}{\Omega}R^3\Omega\frac{\partial \mathbi{l}}{\partial R}\right)\\ \nonumber & &
 +{} \frac{1}{R}\frac{\partial}{\partial R}\left(Q_3\Sigma \frac{c_{\rm s}^2}{\Omega} R^3\Omega \mathbi{l}\times\frac{\partial \mathbi{l}}{\partial R}\right)\,,
\end{eqnarray}
where we have used the relation ${\cal I} = \Sigma c_{\rm s}^2/\Omega^2$ for the azimuthally averaged second vertical moment of the density.\footnote{Strictly speaking the rhs of this relation is missing a factor of order unity dependent on the warp amplitude $\left|\psi\right|$, as the forcing in a warp restricts the hydrostatic balance assumed in its derivation. This affects the viscosity coefficients $Q_1,Q_2,Q_3$ significantly for large $\left|\psi\right|$, and so in this case we must use the form of \cite{Ogilvie2000} taking this effect into account (Ogilvie, private communication).}   We should compare these equations with (\ref{conmass}) \& (\ref{evo2}). The two sets of equations become identical given three relations between the torque coefficients ($\nu_1,\nu_2,\nu_3$) and the effective viscosity coefficients ($Q_1,Q_2,Q_3$) \citep{LP2010,NK2012}, i.e.
\begin{equation}
\nu_1 \propto \alpha_1\left(\alpha,\left|\psi\right|\right) = \frac{\Omega}{R\Omega^\prime}Q_1\left(\alpha,\left|\psi\right|\right) = -\frac{2}{3}Q_1\left(\alpha,\left|\psi\right|\right)\,,
\end{equation}
\begin{equation}
\hspace{-1.27in}\nu_2 \propto \alpha_2\left(\alpha,\left|\psi\right|\right) = 2Q_2\left(\alpha,\left|\psi\right|\right)
\end{equation}
and
\begin{equation}
\hspace{-1.27in}\nu_3 \propto \alpha_3\left(\alpha,\left|\psi\right|\right) = Q_3\left(\alpha,\left|\psi\right|\right)\,,
\end{equation}
where $\left|\psi\right| = R\left|\partial \mathbi{l}/\partial R\right|$ is the warp amplitude and the first relation is given in simplified form for the case of Keplerian rotation. 

The coefficients $Q_1$, $Q_2$ \& $Q_3$ depend on the warp amplitude $\left|\psi\right| = R\left|\partial \mathbi{l}/\partial R\right|$, $\alpha$ and the orbital shear (see Figures 3, 4 \& 5 of \citealt{Ogilvie1999}). In various limits their approximate values can be obtained analytically (e.g. Section 7.3 of \citealt{Ogilvie1999}) but in general need to be computed numerically. For a complete generalisation of the Shakura--Sunyaev theory of planar discs to diffusive warped discs, see \cite{Ogilvie2000} where the effects of viscous dissipation and radiation transport are also included in the viscosity coefficients.

The equations derived in this section, and those by \cite{Ogilvie1999,Ogilvie2000}, hold when $\alpha > H/R$ (damping the otherwise uncontrolled resonant response to the epicyclic forcing) and when $H/R \ll 1$. However, the main uncertainty in modelling a `real' astrophysical disc in this manner is the nature of the turbulence driving angular momentum transport. The approach adopted above is supported by strong evidence (analytical, numerical and observational) but it is certainly not beyond doubt. We comment on this in more detail in Section~\ref{viscosity} below.

\subsection{Waves}
\label{waves}

The evolution equations for a wave--like disc were first derived by \cite{PL1995}, \cite{DI1997} and \cite{LO2000}. All of these papers linearise the fluid equations, and there is very little work on the nonlinear behaviour of wave--like discs. \cite{Ogilvie2006} extends the equations into the weakly nonlinear regime, but this has yet to be followed up in any detail. There is good reason to believe that the nonlinear dynamics may well display many interesting features \citep{Gammieetal2000,OL2013b,OL2013a}, but there remains no fully nonlinear theory for warp propagation in this case. However, the linear dynamics of wave--like discs is reasonably well understood and we discuss this here.

\subsubsection{Evolution equations}

The equations of \cite{LO2000} follow a similar notation to that used in Section~\ref{diffeqns} and so we shall use that here also. The linearised equations of motion for a wave--like disc are one for the conservation of angular momentum
\begin{equation}
\label{LO2000eq1}
\Sigma R^2 \Omega \frac{\partial \mathbi{l}}{\partial t} = \frac{1}{R}\frac{\partial \mathbi{G}}{\partial R} + \mathbi{T}
\end{equation}
and one for the evolution of the internal torque
\begin{equation}
\label{LO2000eq2}
\frac{\partial \mathbi{G}}{\partial t} + \left(\frac{\kappa^2-\Omega^2}{2\Omega}\right)\mathbi{l}\times \mathbi{G} + \alpha\Omega\mathbi{G} = \Sigma R^3 \Omega \frac{c_s^2}{4}\frac{\partial \mathbi{l}}{\partial R}\,.
\end{equation}
The external torque $\mathbi{T}$ can arise from a variety of effects, but is zero if the disc is precisely Keplerian and there are no external effects. The second term on the lhs of (\ref{LO2000eq2}) is also zero when the disc is Keplerian ($\Omega = \Omega_z = \kappa$). The third term on the lhs describes damping of propagating waves, which occurs on a timescale $\sim 1/\left(\alpha\Omega\right)$ -- we shall refer to this as $\alpha$ damping. If we compare (\ref{LO2000eq1}) to (\ref{evo1}) we can see that these equations make the assumption that $\Sigma\left(R\right)$ is independent of time and the radial velocity is zero -- i.e. there is no diffusion of mass or advection of angular momentum in these equations.

In the simplest case of an inviscid ($\alpha=0$) Keplerian disc the equations become
\begin{equation}
\label{wave1}
\Sigma R^2 \Omega \frac{\partial \mathbi{l}}{\partial t} = \frac{1}{R}\frac{\partial \mathbi{G}}{\partial R}
\end{equation}
and
\begin{equation}
\label{wave2}
\frac{\partial \mathbi{G}}{\partial t} = \Sigma R^3 \Omega \frac{c_s^2}{4}\frac{\partial \mathbi{l}}{\partial R}\,,
\end{equation}
which can be combined into a single equation by eliminating $\mathbi{G}$
\begin{equation}
\frac{\partial^2\mathbi{l}}{\partial t^2} = \frac{1}{\Sigma R^3 \Omega} \frac{\partial}{\partial R} \left(\Sigma R^3\Omega \frac{c_{\rm s}^2}{4} \frac{\partial \mathbi{l}}{\partial R}\right)\,.
\end{equation}
Equation~5 of \cite{Ogilvie2006} shows that when $\Sigma H$ is independent of $R$, this can be recast into a classical wave equation for the disc tilt. This shows that the wave speed is $V_{\rm w} = c_{\rm s}/2$ \citep{PL1995}. The wave speed can be evaluated in a similar way from equations A53--A56 of \cite{LO2000} to give the more general result that $V_{\rm w}^2 = {\cal I}\Omega^2/4\Sigma$.

With the equations in this reduced form (\ref{wave1} \& \ref{wave2}) it is easy to see where the terms come from. Equation (\ref{wave1}) is a statement of angular momentum conservation in the case where $\Sigma$ is independent of time and so $V_R = 0$ (cf. Equation~\ref{evo1}). Equation (\ref{wave2}) expresses how the internal torque changes because of the pressure applied by a warp, which might be written as $\partial\mathbi{G}/\partial t = RH.\rho V_{\rm w}^2.R^2\Omega.\partial \mathbi{l}/\partial R$, which with $\rho = \Sigma/H$ gives (\ref{wave2}). This implies that each ring of the disc responds as if hit by a pressure wave with velocity $V_{\rm w}$. With such a simple approach to the internal torque we cannot recover the wave speed, but this allows physical insight into the disc response to a warp. By including a small $\alpha$ damping of the wave propagation, and any non--Keplerian terms due to the potential, we get the full evolution equations derived rigorously by \cite{LO2000}.

Taking the (artificial) limit that $\alpha$ is large in these equations (\ref{LO2000eq2}) becomes
\begin{equation}
\alpha \Omega \mathbi{G} = \Sigma R^3 \Omega \frac{c_{\rm s}^2}{4}\frac{\partial \mathbi{l}}{\partial R}\,.
\end{equation}
Substituting this into (\ref{LO2000eq1}) gives
\begin{equation}
\Sigma R^2\Omega \frac{\partial \mathbi{l}}{\partial t} = \frac{1}{R}\frac{\partial}{\partial R}\left(\Sigma R^3\Omega \frac{c_{\rm s}^2}{4\alpha\Omega}\frac{\partial\mathbi{l}}{\partial R}\right)\,.
\end{equation}
Now we can see that in the linear regime with $\nu_2 = c_{\rm s}H/2\alpha$ this torque is exactly that derived to communicate the disc tilt in the diffusive case. This is not by chance; the physics of the two terms is precisely the same in this approximation. It is likely therefore that for specific (short) timescales one might use the linearised wave equations with strong damping to model a linear diffusive disc \citep[see e.g.][]{Facchinietal2013}, but that this approach is ultimately missing much of the physics in warped discs (mass diffusion, advection of angular momentum and internal precession). Good agreement between the approaches may be found in cases where these effects are minimised by extra physics in the problem, for example where tidal torques inhibit mass flow.

As a warp wave propagates in the disc it can lead to either a local or global bending of the disc. For a single wave the criterion for this is effectively governed by the wavelength ($\lambda$) of the disturbance. For $H \ll \lambda < R$ the disc responds by warping, but when $\lambda > R$ the wave is unable to bend the disc which instead responds by tilting as a whole. This behaviour was observed in the simulations of \cite{NP2010} where a warp was induced at the outer edge of a disc by a passing perturber. This warp had a wavelength of approximately one third of the outer disc radius, leading to warped outer regions. As the warp wave propagates inwards the inner disc ($R < \lambda$) tilts as a whole while remaining planar.

In wave--dominated discs there is also the possibility that a disc might respond to a disturbance by globally precessing. In the paragraph above this clearly occurs when $\lambda > R_{\rm out}$. However, it can also occur when the wave communication across the entire disc is short compared to any precession time of the disc. In this case the disc is able to share the precession generated at each radius across the entire disc and respond as a cohesive whole. This has been observed in many simulations \citep[e.g.][]{LP1997,Fragileetal2007,FN2010}, although there appears to be some ambiguity in the literature as to the exact criteria for the onset of this behaviour. It is likely that it requires waves to damp non--locally (effectively $\alpha \ll H/R$) and that the wave travel time across the disc ($\sim R/c_{\rm s}$) should be shorter than any precession induced in the disc. 

\cite{NK2013} pointed out that when only a small piece of the disc is simulated these conditions can be artificially met, facilitating repeated global precession.  This occurs as the propagating wave reflects off the outer boundary, which in reality should be orders of magnitude further out. This allows unphysical global precession. Instead the wave should leave behind a steady disc shape \citep[e.g.][]{Lubowetal2002}. In other cases, discs may be kept radially narrow by extra physics such as tidal truncation, allowing physical global precession \citep[e.g.][]{LF2013}. Care must also be taken when simulating warped discs, as the propagation of waves depends sensitively on the ratio of $\alpha$ and $H/R$, these parameters must be accurately modelled for the system of interest. Often thick discs are employed in simulations to increase vertical resolution, without observational or theoretical motivation. Similarly the turbulent viscosity may not be modelled correctly if numerical affects (e.g. low resolution) lead to strong numerical dissipation, or if the effective $\alpha$ in simulations which resolve turbulence is smaller than that implied by observations \citep{Kingetal2007}.

The main untouched area for wave--like discs is a thorough understanding of their nonlinear evolution -- \cite{Ogilvie1999} discusses this for the diffusive case. In the wave--like case \cite{Ogilvie2006} extends the equations of \cite{LO2000} to the weakly nonlinear regime, but the predictions made are yet to be tested by a hydrodynamics code. Also, large warps may be hydrodynamically unstable for small $\alpha$. \cite{Gammieetal2000} predict that such warps are subject to a parametric instability which leads to dissipation and enhanced wave damping. More recently \cite{OL2013b,OL2013a} have developed a local model for warped discs and have used this to explore the onset of turbulence in this hydrodynamic instability. Such instabilities have not appeared so far in global simulations, probably because of insufficient resolution and artificial dissipation effects.

\subsection{Viscosity}
\label{viscosity}
We have seen in the previous sections that the usual approach to implementing a viscosity for warped discs is to adopt the $\alpha$ parameterisation, where the local stress is proportional to the pressure. This means that each component of the shear (horizontal and vertical) is damped by viscous dissipation at the same average rates. This is the origin of the term `isotropic viscosity', but this does {\it not} imply that the effective viscosities (torque coefficients $\nu_1$, $\nu_2$ \& $\nu_3$) are isotropic. Instead, the internal structure of a warp means that these torque coefficients take very different values that depend on the disc shear, $\alpha$ and warp amplitude. For small $\alpha$ and small warps, $\nu_2$ is significantly stronger because of the resonance between disc orbits and forcing in a warp --  the resonance is controlled by $\alpha$ damping, so $\nu_2$ is inversely proportional to $\alpha$. The angular momentum transport is mainly through Reynolds stresses rather than viscous stresses, but this process is well described by a viscosity.\footnote{There is a long history, in both the astronomical and fluids literatures, of modelling turbulent fluids with effective viscosities.}

\cite{PP1983} and \cite{Ogilvie1999} assume that $\alpha$ acts isotropically, and this is responsible for the $\alpha_2 \propto 1/\alpha$ relation. However, it is not obvious that this local isotropy holds for a viscosity driven by MHD effects, typically due to the magnetorotational instability (MRI). \cite{Pringle1992} points out that the azimuthal shear is secular, but that the vertical shear is oscillatory. This means that gas parcels displaced by $\Delta R$ drift further and further apart, but those at the same radius, but displaced by $\Delta z$, simply oscillate. In the simplistic MRI picture it is then likely that more energy is dissipated in the azimuthal viscosity, leading to a larger torque. However, reduced dissipation in the vertical direction may lead to an uncontrolled resonant response and thus a larger $\nu_2$ torque. Thus simple arguments leave it entirely unclear how the inclusion of MHD effects should change this picture. 

If we instead assume that the MRI drives turbulence, and that the velocity field is uncorrelated on scales $\ll H$ (as seems probable from numerical simulations, e.g. Fig.~14 of \citealt{Simonetal2012}) then the action of the turbulent viscosity seems likely not to care about the direction of the shear on which it acts. This supports the isotropic $\alpha$ assumption.

The question of whether $\alpha$ is isotropic has been studied with numerical \citep{Torkelssonetal2000} and analytical \citep{Ogilvie2003} techniques. Both of these investigations conclude that the isotropic $\alpha$ assumption is valid for an MRI--turbulent disc. There have also been attempts to interpret the dynamics of warped discs by using observations. These currently support the isotropic picture \citep{Kingetal2013}.

The viscosity assumption in the diffusive and wave--like cases are effectively the same. As shown above, the term which describes $\alpha$ damping of propagating waves in (\ref{LO2000eq2}) can be readily understood as exactly the $\nu_2$ torque in (\ref{pringledLdt}). The physics of both torques is the same -- $\alpha$ damping controls the resonant response to radial pressure gradients which force the gas at the local orbital frequency (when the disc is Keplerian). When $\alpha < H/R$ the disc responds by propagating a wave which damps non--locally through $\alpha$. For $\alpha > H/R$ this can instead be thought of as the wave being damped locally, and thus acting in a diffusive manner.

\section{Conclusions}
Astrophysical discs are often warped. This occurs when an initially planar disc is misaligned to a component of the potential (e.g. the spin of a black hole, or the orbit of a companion star) or when the disc becomes unstable to tilting by e.g. tidal \citep{Lubow1992a} or radiation \citep{Pringle1996} effects. The main extra physics in a warped disc is the inclusion of a second component of the orbital shear which drives an oscillatory pressure gradient in the fluid. In a near--Keplerian disc this results in a resonant response which communicates the tilted component of angular momentum. As this response is damped by an $\alpha$ viscosity, it depends inversely on that viscosity. This behaviour allows the dynamics of warped discs to split broadly into two regimes; one where the warp propagates diffusively and the other where it propagates as waves. 

The equations of motion in the diffusive case are well understood, in both the linear and nonlinear regimes, when the viscous stress is proportional to the local pressure (i.e. isotropic). The evolution equations were first derived by \cite{PP1983} and then extended to the fully nonlinear regime by \cite{Ogilvie1999}. This is generalised even further to include viscous dissipation and radiative transport by \cite{Ogilvie2000}. 

The equations of motion in the wave--like case are only really understood in the linear regime, again with an $\alpha$ damping. The nonlinear case is difficult to explore numerically as it requires modelling the disc with little dissipation, and thus high resolution. However, detailed investigation into this case is likely to be fruitful in revealing complex disc behaviour such as the parametric instability \citep{Gammieetal2000}. This is relevant to protostellar discs where the disc turbulence may be heavily quenched in dead zones \citep{Gammie1996}.

\begin{figure}
  \begin{center}
    \includegraphics[angle=0,width=\columnwidth]{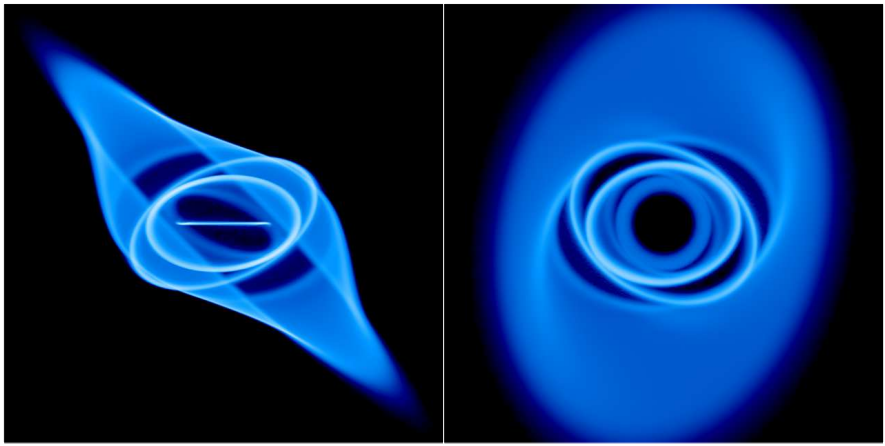}
    \caption{Column density projection of the tearing disc structure in a simulation following the method of \cite{Nixonetal2012b} from \cite{NS2014}.}
    \label{tearing}
  \end{center}
\end{figure}

There are many aspects of warped disc physics which are yet to be fully understood, from the inclusion of MHD, radiation warping and self--gravity to nonlinear hydrodynamics in strong warps. Very recently it has been shown that warped discs are capable of breaking into distinct planes and that this can significantly alter the disc's behaviour \citep{Nixonetal2012b,KN2013}. This evolution (see Fig.~\ref{tearing}) has been observed in a variety of codes, with different numerical methods and different physics \citep{Larwoodetal1996,LP1997,LP2010,FN2010,NK2012,Nixonetal2012b,Nixonetal2013}, so appears to be a generic feature of warped disc behaviour. It has yet to be shown that an MRI turbulent MHD disc is capable of breaking, but there has not yet been any simulation of a setup in which this would have been the likely result.

Warped discs are relevant to a variety of astrophysical systems, including protostars, X--ray binaries, tidal disruption events, AGN and others. Each of these systems have different properties and environments which allow for a wide variety in the disc's response to a warp. Warps are often induced when a disc forms misaligned to symmetry axes of the potential, for example the spin of a black hole or the orbit of a binary. However, even initially planar discs can become warped through a variety of effects, e.g. radiation warping, tidal torques, winds or magnetic fields. Each of these problems and their implications are yet to be fully understood, but with the growing computational power and complexity of physics employed in numerical codes it is likely that many questions will be answered in the next few years. 

\acknowledgement We thank Phil Armitage, Giuseppe Lodato, Steve Lubow, Rebecca Martin, Rebecca Nealon, Daniel Price \& Jim Pringle for providing thoughtful comments on the manuscript. CN thanks NASA for support through the Einstein Fellowship Programme, grant PF2-130098. Astrophysics research at Leicester is supported by an STFC Consolidated Grant.

\bibliographystyle{mn2e}
\bibliography{nixon}

\begin{thebibliography}{58}
\expandafter\ifx\csname natexlab\endcsname\relax\def\natexlab#1{#1}\fi

\bibitem[{{Balbus} \& {Hawley}(1991)}]{BH1991}
{Balbus} S.~A., {Hawley} J.~F., 1991, \apj, 376, 214

\bibitem[{{Bardeen} \& {Petterson}(1975)}]{BP1975}
{Bardeen} J.~M., {Petterson} J.~A., 1975, \apjl, 195, L65+

\bibitem[{{Demianski} \& {Ivanov}(1997)}]{DI1997}
{Demianski} M., {Ivanov} P.~B., 1997, \aap, 324, 829

\bibitem[{{Facchini} {et~al}\mbox{.}(2013){Facchini}, {Lodato}, \&
  {Price}}]{Facchinietal2013}
{Facchini} S. {et~al.}, 2013, \mnras, 433, 2142

\bibitem[{{Fleming} \& {Stone}(2003)}]{FS2003}
{Fleming} T., {Stone} J.~M., 2003, \apj, 585, 908

\bibitem[{{Fragile} {et~al}\mbox{.}(2007){Fragile}, {Blaes}, {Anninos}, \&
  {Salmonson}}]{Fragileetal2007}
{Fragile} P.~C. {et~al.}, 2007, \apj, 668, 417

\bibitem[{{Fragner} \& {Nelson}(2010)}]{FN2010}
{Fragner} M.~M., {Nelson} R.~P., 2010, \aap, 511, A77

\bibitem[{{Frank} {et~al}\mbox{.}(2002){Frank}, {King}, \&
  {Raine}}]{Franketal2002}
{Frank} J. {et~al.}, 2002, {Accretion Power in Astrophysics}

\bibitem[{{Gammie}(1996)}]{Gammie1996}
{Gammie} C.~F., 1996, \apj, 457, 355

\bibitem[{{Gammie} {et~al}\mbox{.}(2000){Gammie}, {Goodman}, \&
  {Ogilvie}}]{Gammieetal2000}
{Gammie} C.~F. {et~al.}, 2000, \mnras, 318, 1005

\bibitem[{{Gressel} {et~al}\mbox{.}(2012){Gressel}, {Nelson}, \&
  {Turner}}]{Gresseletal2012}
{Gressel} O. {et~al.}, 2012, \mnras, 422, 1140

\bibitem[{{Hartmann} {et~al}\mbox{.}(1998){Hartmann}, {Calvet}, {Gullbring}, \&
  {D'Alessio}}]{Hartmannetal1998}
{Hartmann} L. {et~al.}, 1998, \apj, 495, 385

\bibitem[{{Hatchett} {et~al}\mbox{.}(1981){Hatchett}, {Begelman}, \&
  {Sarazin}}]{Hatchettetal1981}
{Hatchett} S.~P. {et~al.}, 1981, \apj, 247, 677

\bibitem[{{King} \& {Nixon}(2013)}]{KN2013}
{King} A., {Nixon} C., 2013, Classical and Quantum Gravity, 30, 244006

\bibitem[{{King} {et~al}\mbox{.}(2013){King}, {Livio}, {Lubow}, \&
  {Pringle}}]{Kingetal2013}
{King} A.~R. {et~al.}, 2013, \mnras, 431, 2655

\bibitem[{{King} {et~al}\mbox{.}(2007){King}, {Pringle}, \&
  {Livio}}]{Kingetal2007}
{King} A.~R. {et~al.}, 2007, \mnras, 376, 1740

\bibitem[{{Korycansky} \& {Pringle}(1995)}]{KP1995}
{Korycansky} D.~G., {Pringle} J.~E., 1995, \mnras, 272, 618

\bibitem[{{Kumar} \& {Pringle}(1985)}]{KP1985}
{Kumar} S., {Pringle} J.~E., 1985, \mnras, 213, 435

\bibitem[{{Lai}(1999)}]{Lai1999}
{Lai} D., 1999, \apj, 524, 1030

\bibitem[{{Larwood} {et~al}\mbox{.}(1996){Larwood}, {Nelson}, {Papaloizou}, \&
  {Terquem}}]{Larwoodetal1996}
{Larwood} J.~D. {et~al.}, 1996, \mnras, 282, 597

\bibitem[{{Larwood} \& {Papaloizou}(1997)}]{LP1997}
{Larwood} J.~D., {Papaloizou} J.~C.~B., 1997, \mnras, 285, 288

\bibitem[{{Lodato} \& {Facchini}(2013)}]{LF2013}
{Lodato} G., {Facchini} S., 2013, \mnras, 433, 2157

\bibitem[{{Lodato} \& {Price}(2010)}]{LP2010}
{Lodato} G., {Price} D.~J., 2010, \mnras, 405, 1212

\bibitem[{{Lodato} \& {Pringle}(2006)}]{LP2006}
{Lodato} G., {Pringle} J.~E., 2006, \mnras, 368, 1196

\bibitem[{{Lodato} \& {Pringle}(2007)}]{LP2007}
{Lodato} G., {Pringle} J.~E., 2007, \mnras, 381, 1287

\bibitem[{{Lubow}(1992)}]{Lubow1992a}
{Lubow} S.~H., 1992, \apj, 398, 525

\bibitem[{{Lubow} \& {Ogilvie}(2000)}]{LO2000}
{Lubow} S.~H., {Ogilvie} G.~I., 2000, \apj, 538, 326

\bibitem[{{Lubow} {et~al}\mbox{.}(2002){Lubow}, {Ogilvie}, \&
  {Pringle}}]{Lubowetal2002}
{Lubow} S.~H. {et~al.}, 2002, \mnras, 337, 706

\bibitem[{{Lubow} \& {Pringle}(1993)}]{LP1993}
{Lubow} S.~H., {Pringle} J.~E., 1993, \apj, 409, 360

\bibitem[{{Martin} \& {Lubow}(2014)}]{ML2014}
{Martin} R.~G., {Lubow} S.~H., 2014, \mnras, 437, 682

\bibitem[{{Nixon} \& {King}(2013)}]{NK2013}
{Nixon} C., {King} A., 2013, \apjl, 765, L7

\bibitem[{{Nixon} {et~al}\mbox{.}(2013){Nixon}, {King}, \&
  {Price}}]{Nixonetal2013}
{Nixon} C. {et~al.}, 2013, \mnras, 434, 1946

\bibitem[{{Nixon} {et~al}\mbox{.}(2012){Nixon}, {King}, {Price}, \&
  {Frank}}]{Nixonetal2012b}
{Nixon} C. {et~al.}, 2012, \apjl, 757, L24

\bibitem[{{Nixon} \& {Salvesen}(2014)}]{NS2014}
{Nixon} C., {Salvesen} G., 2014, \mnras, 437, 3994

\bibitem[{{Nixon} \& {King}(2012)}]{NK2012}
{Nixon} C.~J., {King} A.~R., 2012, \mnras, 421, 1201

\bibitem[{{Nixon} \& {Pringle}(2010)}]{NP2010}
{Nixon} C.~J., {Pringle} J.~E., 2010, \mnras, 403, 1887

\bibitem[{{Ogilvie}(1999)}]{Ogilvie1999}
{Ogilvie} G.~I., 1999, \mnras, 304, 557

\bibitem[{{Ogilvie}(2000)}]{Ogilvie2000}
{Ogilvie} G.~I., 2000, \mnras, 317, 607

\bibitem[{{Ogilvie}(2003)}]{Ogilvie2003}
{Ogilvie} G.~I., 2003, \mnras, 340, 969

\bibitem[{{Ogilvie}(2006)}]{Ogilvie2006}
{Ogilvie} G.~I., 2006, \mnras, 365, 977

\bibitem[{{Ogilvie} \& {Latter}(2013{\natexlab{a}})}]{OL2013b}
{Ogilvie} G.~I., {Latter} H.~N., 2013{\natexlab{a}}, \mnras, 433, 2420

\bibitem[{{Ogilvie} \& {Latter}(2013{\natexlab{b}})}]{OL2013a}
{Ogilvie} G.~I., {Latter} H.~N., 2013{\natexlab{b}}, \mnras, 433, 2403

\bibitem[{{Papaloizou} \& {Lin}(1995)}]{PL1995}
{Papaloizou} J.~C.~B., {Lin} D.~N.~C., 1995, \apj, 438, 841

\bibitem[{{Papaloizou} \& {Pringle}(1983)}]{PP1983}
{Papaloizou} J.~C.~B., {Pringle} J.~E., 1983, \mnras, 202, 1181

\bibitem[{{Papaloizou} \& {Terquem}(1995)}]{PT1995}
{Papaloizou} J.~C.~B., {Terquem} C., 1995, \mnras, 274, 987

\bibitem[{{Petterson}(1977)}]{Petterson1977a}
{Petterson} J.~A., 1977, \apj, 214, 550

\bibitem[{{Petterson}(1978)}]{Petterson1978}
{Petterson} J.~A., 1978, \apj, 226, 253

\bibitem[{{Pringle}(1981)}]{Pringle1981}
{Pringle} J.~E., 1981, \araa, 19, 137

\bibitem[{{Pringle}(1992)}]{Pringle1992}
{Pringle} J.~E., 1992, \mnras, 258, 811

\bibitem[{{Pringle}(1996)}]{Pringle1996}
{Pringle} J.~E., 1996, \mnras, 281, 357

\bibitem[{{Pringle}(1997)}]{Pringle1997}
{Pringle} J.~E., 1997, \mnras, 292, 136

\bibitem[{{Pringle}(1999)}]{Pringle1999}
{Pringle} J.~E., 1999, ASPCS, 160, 53

\bibitem[{{Pringle} \& {Rees}(1972)}]{PR1972}
{Pringle} J.~E., {Rees} M.~J., 1972, \aap, 21, 1

\bibitem[{{Shakura} \& {Sunyaev}(1973)}]{SS1973}
{Shakura} N.~I., {Sunyaev} R.~A., 1973, \aap, 24, 337

\bibitem[{{Simon} {et~al}\mbox{.}(2011){Simon}, {Armitage}, \&
  {Beckwith}}]{Simonetal2011}
{Simon} J.~B. {et~al.}, 2011, \apj, 743, 17

\bibitem[{{Simon} {et~al}\mbox{.}(2012){Simon}, {Beckwith}, \&
  {Armitage}}]{Simonetal2012}
{Simon} J.~B. {et~al.}, 2012, \mnras, 422, 2685

\bibitem[{{Torkelsson} {et~al}\mbox{.}(2000){Torkelsson}, {Ogilvie},
  {Brandenburg}, {Pringle}, {Nordlund}, \& {Stein}}]{Torkelssonetal2000}
{Torkelsson} U. {et~al.}, 2000, \mnras, 318, 47

\bibitem[{{Wijers} \& {Pringle}(1999)}]{WP1999}
{Wijers} R.~A.~M.~J., {Pringle} J.~E., 1999, \mnras, 308, 207

\end{thebibliography}

\end{document}